# Spontaneous entropy decrease and its statistical formula

## Xing Xiu-San

(Department of Physics,Beijing Institute of Technology,Beijing 100081,China)

email: xingxs@sohu.com

**Abstract**

Why can the world resist the law of entropy increase and produce self-organizing structure? Does the entropy of an isolated system always only increase and never decrease? Can be thermodymamic degradation and self-organizing evolution united? How to unite? In this paper starting out from nonequilibrium entropy evolution equation we proved that a new entropy decrease could spontaneously emerge in nonequilibrium system with internal attractive interaction. This new entropy decrease coexists with the traditional law of entropy increase, both of them countervail each other, so that the total entropy of isolated system can be able to decrease. It not only makes isolated system but also helps open system to produce self-organizing structure. We first derived a statistical formula for this new entropy decrease rate, and compared it both in mathematical form and in microscopic physical foundation with the statistical formula for the law of entropy increase which was derived by us some years ago. Furthermore, we gave the formulas for the time rate of change of total entropy in isolated system and open system. The former is equal to the sum of the formula for the law of entropy increase and the formula for the new entropy decrease rate, the latter is the algebraic sum of the formulas for entropy increase, entropy decrease and entropy flow. All of them manifest the unity of thermodynamic degradation and self-organizing evolution. As the application of the new theoretical formulas, we discussed qualitatively the emergency of inhomogeneous structure in two real isolated systems including clarifying the inference about the heat death of the universe.

**1. Introduction**

The law of entropy increase[1-4], i.e. the second law of thermodynamics expressed by the entropy, is a fundamental law in nature. It shows if an isolated system is not in a statistical equilibrium state, its macroscopic entropy will increase with time, until ultimately the system reaches a complete equilibrium state where the entropy attains its maximum value. According to the inference of this law, the universe is as isolated system, it also ought to degrade into a complete statistical equilibrium state, i.e. the so-called heat death state[3-4]. Then, the entropy and randomness in the universe are at their maximum, there are only gas molecules with homogeneous distribution, all macroscopic mechanical energy degrades into heat energy of gas molecules, no further change occurs. However, the real world is another scene. Everywhere there is order and structure: stars, galaxies, plants and animals etc. They are always incessantly evolving. When the law of entropy increase occupies a dominant position, why can an isolated system create order structure? Why can the life from nothing to some thing with simple atoms and molecules organize itself into a whole? Why is the universe still able to bring forth stars, galaxies and does not stop at or dissolve into simple gas? Why can they resist the law of entropy increase and produce self-organizing structure? Whether or not because they are also governed by a power of some unknown entropy decrease inherent in the system? If so, how does this power of entropy decrease coexist and countervail with the law of entropy increase, and hence both of them form the unity of opposites between thermodynamic degradation and self-organizing evolution? What is its dynamical mechanism? And what is its mathematical expression? What is the difference in mathematical form and microscopic physical foundation between the formulas for this entropy decrease and the law of entropy increase? Under what condition can an isolated system overcome the law of entropy increase and produce self-organizing structure? And what is the difference in entropy condition for growing self-organizing structure between open system and isolated system? Can all these problems be solved from a nonequilibrium entropy evolution equation in a unified fashion?

In the late half of the twentieth century, the publications of the theory of dissipative structures[5], synergetics[6] and the hypercycle[7] marked an important progress of quantitative theories in self-organization. However, these theories, including the formal entropy theory decomposing the entropy change into the sum of the entropy flow and the entropy production, discuss only the problems of open system but not isolated system. From the point of view of exploring that what system can spontaneously produce a entropy decrease to be a match for the law of entropy increase, they all have no relation. Of course, they also have no help to clarify the puzzle of the heat death. In recent years during doing research on the fundamental problems of nonequilibrium statistical physics, we[8-12] proposed a new equation of time-reveral asymmetry in place of the Liouville equation of time-reversal symmetry as the fundamental equation of nonequilibrium statistical physics. That is the anomalous Langevin equation in 6N dimensional phase space or its equivalent Liouville diffusion equation. Starting from this equation we decided succinctly the hydrodynamic equations such as diffusion equation, thermal conductivity equation and Navier-Stokes equation. Furthermore we presented a nonlinear evolution equation of Gibbs and Boltzmann nonequilibrium entropy density changing with time-space (called for short nonequilibrium entropy evolution equation), predicted the existence of entropy diffusion, and obtained a concise statistical formula for the law of entropy increase. In this paper we solved all above mentioned problems on the basis of these new known works, especially we first proved that a new entropy decrease can spontaneously emerge in nonequilibrium system with internal attractive interaction, and derived its statistical formula. Just that the power of this new entropy decrease countervails the law of entropy increase leads to the total entropy of an isolated system to be able to decrease, and not only makes isolated system but also helps open system to produce self-organizing structure.

For the sake of saving space of the paper, the following quantitative expressions are limited in 6 dimensional phase space.

## 2. Nonequilibrium entropy evolution equation

According to nonequilibrium statistical physics, nonequilibrium entropy in 6 dimensional phase space can be defined as[1,13,14]

$$S(t) = -k \int f_1(\boldsymbol{x},t) \ln \frac{f_1(\boldsymbol{x},t)}{f_{10}(\boldsymbol{x})} d\boldsymbol{x} + S_0 = \int S_{vp} d\boldsymbol{x} + S_0 \qquad (1)$$

or $$S(t) = -k \int f_1(\boldsymbol{x},t) \ln f_1(\boldsymbol{x},t) d\boldsymbol{x} = \int S_{vp} d\boldsymbol{x} \qquad (1a)$$

where $x = (q, p)$ is the state vector in 6 dimensional phase space, $q$ and $p$ are the position and momentum of the single particle, $f_1(x,t) = f_1(q,p,t)$ is the single particle probability density at time $t$, $f_{10}(x)$ is its equilibrium probability density. $S_0$ is equilibrium entropy in 6 dimensional phase space, $k$ is the Boltzmanm constant, $S_{vp}$ is entropy density in 6 dimensional phase space. The reason why we used the expression (1) but not(1a) as the definition of nonequilibrium entropy will explain in the following section 2.

Just the same as that the BBGKY equation hierarchy[14,15] can be derived from the Liouville equation, the BBGKY diffusion equation[8-11] of the single-particle probability density can also be derived from the Liouville diffusion equation as follows

$$\left[\frac{\partial}{\partial t} + \frac{p}{m} \times \nabla_q + F \times \nabla_p \right] f_1(x,t) = N \int (\nabla_q f) \times \nabla_p f_2(x, x_1, t) dx_1 + D \nabla_q^2 f_1(x,t) \quad (2)$$

Where $f_2(x, x_1, t)$ is two particle joint probability density at time $t$, $F$ is external force, $N$ is particle number in the system, $D$ is self-diffusion coefficient of the particle, $f = f(|q - q_1|)$ is the two-particle interaction potential. The first term on the right-hand side of equation (2) is commonly referred to as the collision integral, now we call it interaction term. We shall see below that the new entropy decrease of an isolated system with internal attractive interaction just is emerged from this term. The second term on the right-hand side of equation (2) comes from the diffusion term of the Liouville diffusion equation, when there is no this term, equation (2) is just the BBGKY equation of single-particle probability density.

Differentiating both side of formula (1) with respect to time $t$ and using the BBCKY diffusion equation (2)(now assume $F = 0$), we obtained nonequilibrium entropy evolution equation in 6 dimensional phase space as follows [9-12]

$$\frac{\partial S_{vp}}{\partial t} = -\nabla_q \times (v S_{vp}) + D \nabla_q^2 S_{vp} + \frac{D}{k f_1}\left[(\nabla_q \ln f_1) S_{vp} - \nabla_q S_{vp}\right]^2 + I(f) \quad (3)$$

where $v = p/m$ is the particle velocity. Equation (3) shows that the time rate of change of nonequilibrium entropy density (on the left-hand side) originates together from its drift in space (the first term on the right-hand side), diffusion (the second term on the right-hand side), increase(the third term on the right-hand side) and the entropy density change contributed by interaction potential energy(the fourth term on the right-hand side)(the discussion on physical meaning of this equation is remained in the end of this letter). Both entropy diffusion and entropy increase come from the diffusion term, the second term on the right–hand side of the BBGKY diffusion equation (2), whose microscopic mechanism is stochastic motion of the particle. The entropy density change rate $I(f)$ contributed by the interaction potential between microscopic particles originates from the first term on the right-hand side of the BBGKY diffusion equation (2), which is the core of this paper.

In order to investigate $l(f)$ and its countervailing with the law of entropy increase, at first let us consider the formula for the law of entropy increase. Integrating the third term on the right-hand side of entropy evolution equation (3) with respect to 6 dimensional phase space, we obtained the formula for entropy increase rate of nonequilibrium system as follows[8-12]

$$P(t) = kD \int f_1(\boldsymbol{x},t) \left[ \nabla_q \ln \frac{f_1(\boldsymbol{x},t)}{f_{10}(\boldsymbol{x})} \right]^2 d\boldsymbol{x} = kD \overline{(\nabla_q q_1)^2} \geq 0 \qquad (4)$$

Where $q_1 = \ln \dfrac{f_1}{f_{10}} = -\ln \dfrac{w_1}{w_{10}} \approx -\dfrac{\Delta w_{10}}{w_{10}}$ can be defined as the percentage departure from equilibrium of the number density of micro-states of the nonequilibrium system in 6 dimensional phase space ($w_1$ and $w_{10}$ for nonequilibrium and equilibrium states), or called the percentage departure from equilibrium for brevity. It can play a role of an independent physical parameter to describe quantitatively how far a nonequilibrium system is from equilibrium as if that the strain or elongation percentage $\in = \ln \dfrac{l}{l_0}$ describes the deformation of solid materials in solid mechanics.

As explained at the beginning of this paper, although people know long ago that the law of entropy increase is a fundamental law in nature, however, what is its microscopic physical basis? Which physical parameter and how does it change with? Can it be described by a quantitative concise formula? This is all along an important problem to be solved in statistical thermodynamics especially in nonequilibrium statistical physics. Formula (4) is the concise statistical formula for entropy increase rate in 6 dimensional phase space derived by us. This is also the quantitative concise statistical formula for the law of entropy increase. It shows that the entropy increase rate $P(t)$ equals to the product of diffusion coefficient $D$, the average value of the square of space gradient of the percentage departure from equilibrium $\overline{(\tilde{N}_q q)^2}$ and Boltzmann constant $k$. It can be seen that the macroscopic entropy increase in nonequilibrium system is caused by spatially stochastic and inhomogeneous departure from the equilibrium of the number density of micro-states. Obviously for nonequilibrium $(q_1 \neq 0)$, spatially inhomogeneous system ($\tilde{N}_q q_1 \neq 0$) with stochastic diffusion $D \neq 0$, the entropy always increases $(P > 0)$. Conversely for equilibrium system $(q_1 = 0)$ or that nonequilibrium but spatially homogeneous system ($\tilde{N}_q q_1 = 0$) or that system only with deterministic but no stochastic motion $(D \neq 0)$, there all are no entropy increase $(P = 0)$.

**3. Formula for entropy decrease rate**

Now let us discuss the core topic: a new entropy decrease can spontaneously emerge in nonequilibrium system with internal attractive interaction. Before proving this proposition, let us consider the relation among a famous kinetic equation — Boltzmann equation, Landau equation and BBGKY diffusion equation (2). In fact the former two is a variety of the latter (when there is no diffusion term) and is applied to describe the motion of a dilute and weakly coupled gas system[14,15] with internal short-range repulsive interaction or collision. More concretely saying, that is to change the first term with $f_2(\boldsymbol{x},\boldsymbol{x}_1,t)$, the two particle interaction term, on the right hand side of equation (2) into collision term only with $f_1(\boldsymbol{x},t)$ of dilate and weakly coupled gas particle. All other terms do not change. As a result, entropy of this nonequilibrium gas system increases due to the existence of short-range repulsive interaction or collision. In other words, entropy increase in the system described by Boltzmann equation and Landau equation originates from short-range repulsive interaction or collision[14,15] between internal particles. This enlights us：when the interaction force among internal particles is attractive but not repulsive, a new entropy decrease should spontaneously emerge in nonequilibrium system. In view of that all closed kinetic equations should be changed from the unclosed BBGKY equation (2), and up to now we did not know what closed kinetic equation should be applied to describe a nonequilibrium statistical system with internal attractive interaction, so the kinetic equation to describe galactic dynamics is yet BBGKY equation [16]. This is the reason why our discussion is based on BBGKY diffusion equation (2). Thus, our proof reduces to: when the internal two-particle interaction potential $f < 0$, the time change rate of entropy $R(t)$ contributed by it and described by the fourth term on the right-hand side of equation(3) should be negative. That is $R(t) = \int I(f)d\boldsymbol{x} < 0$. After some operation, we obtained its mathematical expression

$$R(t) = \int I(f)d\boldsymbol{x}$$

$$= Nk\int (\nabla_q f) \times \left\{ \frac{f_1(\boldsymbol{x},t)}{f_{10}(\boldsymbol{x})} \nabla_p f_{20}(\boldsymbol{x},\boldsymbol{x}_1) - \nabla_p f_2(\boldsymbol{x},\boldsymbol{x}_1,t)\left[1 + \ln\frac{f_1(\boldsymbol{x},t)}{f_{10}(\boldsymbol{x})}\right]\right\} d\boldsymbol{x}d\boldsymbol{x}_1$$

$$= Nk\int f_2(\boldsymbol{x},\boldsymbol{x}_1,t)(\nabla_q f) \times \left\{\nabla_p\left[\ln\frac{f_1(\boldsymbol{x},t)}{f_{10}(\boldsymbol{x})}\right] - \frac{f_{20}(\boldsymbol{x},\boldsymbol{x}_1)}{f_2(\boldsymbol{x},\boldsymbol{x}_1,t)}\nabla_p\left[\frac{f_1(\boldsymbol{x},t)}{f_{10}(\boldsymbol{x})}\right]\right\}d\boldsymbol{x}d\boldsymbol{x}_1 \qquad (5)$$

As mentioned above, since this expression is too complicated to be understood clearly its physical meaning, previously we paid no attention to and wrote it as $-\tilde{N}_q \times \boldsymbol{J}_{vp}$ all along in nonequilibrium entropy evolution equation(3). Now let us simplify expression (5) and investigate its physical meaning. Similar to introducing $q_1$ into formula (4), here we also define

$$q_2 = \ln\frac{f_2}{f_{20}} = -\ln\frac{w_2}{w_{20}}$$ as the percentage departure from equilibrium of nonequilibrium system

in 12 dimensional phase space. Substituting $q_1$ and $q_2$ into last line of expression (5) and through

some operation, we have

$$R(t) = Nk \int f_2(\boldsymbol{x},\boldsymbol{x}_1,t)(\nabla_q f) \times (\nabla_p q_1)\left[1 - e^{-(q_2-q_1)}\right] d\boldsymbol{x} d\boldsymbol{x}_1$$

$$= Nk \overline{(\nabla_q f) \times (\nabla_p q_1)\left[1 - e^{-(q_2-q_1)}\right]} \qquad (6)$$

This is the statistical formula for the time change rate of macroscopic entropy in nonequilibrium system caused by the interaction potential between internal microscopic particles. It shows that the macroscopic entropy change rate $R(t)$ of nonequilibrium system equals to the average value of the product of interaction force $-\tilde{\nabla}_q f$, momentum space gradient $\tilde{\nabla}_q q_1$ of the percentage departure from equilibrium and the difference of the percentage departure from equilibrium of the system between 12 and 6 dimensional phase space $\left[1 - e^{-(q_2-q_1)}\right] \approx q_2 - q_1$ and then all multiplied by $N$ times Boltzmann constant $k$. $R(t)$) is negative or positive, i.e. that the entropy emerges in the system due to interaction potential energy is decreasing or increasing, it is determined by that potential energy is negative or positive. Now let us give this conclusion the following proof. Because $f_1(\boldsymbol{x},t)$ and $f_{10}(\boldsymbol{x})$ are probability density and all are greater than zero, and their order of magnitude are the same, so $f_1/f_{10}$ is greater one order of magnitude than $ln(f_1/f_{10})$. Similarly, because $f_2(\boldsymbol{x},\boldsymbol{x}_1,t)$ is the joint probability density and greater than zero, and satisfied normalization condition $\int f_2(\boldsymbol{x},\boldsymbol{x}_1,t)d\boldsymbol{x}d\boldsymbol{x}_1 = 1$, so $\tilde{\nabla}_p f_2(\boldsymbol{x},\boldsymbol{x}_1,t)$ is a term with minus sign and $\int \tilde{\nabla}_p f_2(\boldsymbol{x},\boldsymbol{x}_1,t)d\boldsymbol{p} = 0$, hence $\tilde{\nabla}_p f_2(\boldsymbol{x},\boldsymbol{x}_1,t)$ may be regarded as a minus sign term. Besides, $\int (\tilde{\nabla}_q f) \times \tilde{\nabla}_p f_2(\boldsymbol{x},\boldsymbol{x}_1,t)d\boldsymbol{p} = 0$, and $\tilde{\nabla}_p f_2(\boldsymbol{x},\boldsymbol{x}_1,t)$ is also the same order of magnitude as $\tilde{\nabla}_p f_{20}(\boldsymbol{x},\boldsymbol{x}_1)$, that is $\tilde{\nabla}_p f_2(\boldsymbol{x},\boldsymbol{x}_1,t) \approx \tilde{\nabla}_p f_{20}(\boldsymbol{x},\boldsymbol{x}_1)$. Substituting this result into the second line from the bottom of formula (5) we obtained

$$R(t) \approx Nk \int (\nabla_q f) \times \left[\frac{f_1(\boldsymbol{x},t)}{f_{10}(\boldsymbol{x})} - \ln \frac{f_1(\boldsymbol{x},t)}{f_{10}(\boldsymbol{x})}\right] \nabla_p f_2(\boldsymbol{x},\boldsymbol{x}_1,t) d\boldsymbol{x} d\boldsymbol{x}_1$$

Under integral of the above formula, the second factor $\left[f_1/f_{10} - \ln(f_1/f_{10})\right]$ is plus and the third factor $\tilde{\nabla}_p f_2(\boldsymbol{x},\boldsymbol{x}_1,t)$ is minus, hence whether $R(t)$ is negative or positive is determined by that the first factor $\tilde{\nabla}_q f$ is plus or minus. In short, when $\tilde{\nabla}_q f < 0$ (repulsive force), $R(t) > 0$; conversely, when $\tilde{\nabla}_q f > 0$ (attractive force), $R(t) < 0$ [the prove of this conclusion obtaining from the following expression (6a) is same but more simple]. In fact, as mentioned above, since the interaction force between dilute and weakly coupled gas atoms is short-range repulsive ($f > 0, \tilde{\nabla}_q f < 0$), the entropy of the system described by Boltzmann equation and Landau equation increases, i.e. $R(t) > 0$. It can be seen from above discussion: when the interaction force between internal particles is repulsive, the entropy of the system increases; conversely, when the interaction force between internal particles is attractive, the entropy of the system decreases. In other words, a new entropy decrease can spontaneously emerge in nonequilibrium system with internal attractive interaction, formula (6) is its quantitative expression.

Now let us consider the physical basis of a new macroscopic entropy decrease spontaneously emerging in statistical thermodynamic system from formula (6) all-sidedly.

1. The system is nonequilibrium, i.e. $q_1 \neq 0$, $q_2 \neq 0$. Conversely, when the system is equilibrium, i.e. $q_1 = 0, q_2 = 0$, then $R = 0$. It should be pointed out here that entropy decrease $R = 0$ and entropy increase $P = 0$ at equilibrium state are direct conclusions of formulas (6) and (4), it need not any additive condition. Compared with formula (4) and (6), if expression (1a) is applied as the definition of nonequilibrium entropy, the total form of nonequilibrium entropy evolution equation (3) do not change, but the third term and the fourth term on the right-hand side of equation, i.e. the real expressions of the entropy increase rate and the entropy decrease rate of the system change into

$$P'(t) = kD \int f_1(\boldsymbol{x},t)\left[\nabla_q \ln f_1(\boldsymbol{x},t)\right]^2 d\boldsymbol{x} \geq 0 \quad (4a)$$

$$R'(t) = Nk \int f_2(\boldsymbol{x},\boldsymbol{x}_1,t)(\nabla_q f) \times \nabla_p \ln f_1(\boldsymbol{x},t) d\boldsymbol{x} d\boldsymbol{x}_1 \quad (6a)$$

Here if we want that $P' = 0$ and $R' = 0$ at equilibrium state, it need $\tilde{\nabla}_q f_1 = 0, \tilde{\nabla}_p f_1 = 0$. In general, this is not reasonable. This is the reason why expression (1) but not expression (1a) is applied as the definition of the nonequilibrium entropy.

2. The departure from equilibrium of the system is inhomogeneous in momentum subspace, i.e. $\tilde{\nabla}_p \theta_1 \neq 0$. Conversely, when the departure from equilibrium of the system is homogeneous, i.e. $\tilde{\nabla}_p \theta_1 = 0$, then $R = 0$.

3. The system is nonlinear. Because according to correlation theory,[15] $f_2(\boldsymbol{x},\boldsymbol{x}_1,t) = f_1(\boldsymbol{x},t)f_1(\boldsymbol{x}_1,t) + g_2(\boldsymbol{x},\boldsymbol{x}_1,t)$, where $g_2(\boldsymbol{x},\boldsymbol{x}_1,\boldsymbol{t})$ is two-particle correlation function. Even in the most simple case, i.e. the system is in statistically independent state, $g_2(\boldsymbol{x},\boldsymbol{x}_1,\boldsymbol{t}) = 0, f_2(\boldsymbol{x},\boldsymbol{x}_1,t) = f_1(\boldsymbol{x},t)f_1(\boldsymbol{x}_1,t), R(t)$ is nonlinear. When the system is in a correlated state, $R(t)$ will be more complicated. If $f_2(\boldsymbol{x},\boldsymbol{x}_1,t) = 0$, the system is linear, $R$ is also equal to zero.

4. The system is in a correlated system, i.e. $f_2(\mathbf{x}, \mathbf{x}_1, t) \neq f_1(\mathbf{x}, t) f_1(\mathbf{x}_1, t)$. Conversely, when the system is in statistically independent state, i.e. $f_2(\mathbf{x}, \mathbf{x}_1, t) = f_1(\mathbf{x}, t) f_1(\mathbf{x}_1, t)$, then $R = 0$. Vlasov equation just describes such a system which has neither entropy increase nor entropy decrease.

5. There is attractive interaction between particles inside the system, i.e. $f < 0$, $-\tilde{\nabla}_q f < 0$, $\tilde{\nabla}_q f > 0$. Conversely, when $f = 0$, then $R = 0$. Thus it can be seen that formula (6) may be regarded as a relational expression between the new macroscopic entropy decrease rate spontaneously emerging in nonequilibrium system and the attractive interaction force between microscopic particles. In other words, if a macroscopic entropy decrease could emerge in nonequilibrium system, then its microscopic physical basis is that there is attractive interaction force between microscopic particles. When the spontaneous entropy decrease rate is $R(t)$, then it need an attractive interaction force $-\tilde{\nabla}_q f < 0$ between microscopic particles.

Here, a relevant fundamental question naturally arise: what is actually the microscopic physical mechanism for the traditional law of entropy increase? According to the result of Boltzmann equation and Landau equation, entropy increase of the system originates from the short-range repulsive interaction or collision[14,15]. Making an inference from this result, if the two-particle interaction force is attractive but not repulsive, the entropy of the system should decrease but not increase, then thermodynamic second law described by the entropy becomes invalid. However, according to formula (4) derived by us, the stochastic diffusion motion of the particle inside the system and inhomogeneous departure from equilibrium of the number density of micro-state is the microscopic physical basis of the macroscopic entropy increase. In other words, the traditional law of entropy increase is caused by the stochastic motion of the particle, but has no direct relation to the interaction force between particles. Because the stochastic motion is inherent in the microscopic particle of statistical thermodynamic system, the law of entropy increase has universal meaning. Only when there is attractive interaction force, a new entropy decrease also emerges in the system except the law of entropy increase. Both of them coexist in a same system and compete with each other.

**4. Unity of thermodynamic degradation and self-organizing evolution**

The evolution in nature has two directions. One is thermodynamic degradation described by thermodynamic second law, which is degenerative evolution of the system spontaneously tending to the direction increasing the degree of disorder. The other is self-organizing evolution whose type is biological evolution, which is newborn evolution of the system spontaneously tending to the direction increasing the degree of order. How can we unite thermodynamic degradation and self-organizing evolution? This is all along a fundamental problem to be solved in nonequilibrium statistical physics. Now let us give the time rate of change of total entropy in an isolated system and open system, and from these obtained formulas discuss this problem.

At first we present a formula for total entropy change rate of an isolated system with internal attractive interaction. Integrating both side of nonequilibrium entropy evolution equation (3) over 6 dimensional phase space and substituting formulas (4) and (6) into this equation, we obtain the formula for the total entropy change rate

$$\frac{\partial_i S}{\partial t} = kD\overline{(\nabla_q q_1)^2} + Nk\overline{(\nabla_q f) \times (\nabla_p q_1)\left[1 - e^{-(q_2 - q_1)}\right]} \qquad (7)$$

The first and second terms on the right –hand side of evolution equation (3) have disappeared from formula (7), because there is no entropy inflow or outflow for an isolated system, and the entropy diffusion only affects the local distribution of entropy density but not affects the total entropy increase or decrease of the system. Formula (7) shows that the total entropy change rate of an isolated system (on the left-hand side) is equal to the sum of the formula for the traditional law of entropy increase (the first term on the right-hand side) and the formula for the new entropy decrease rate (the second term on the right-hand side). The former is positive and originates from the stochastic motion of microscopic particle inside the system, the latter is negative and comes from the attractive interaction between microscopic particles inside the system. Both of them coexist and countervail each other. Entropy increase destroys the order structure and is thermodynamic degradation, entropy decrease produces the order structure and is self-organizing evolution. It is obvious from formula (7), if there is only the first term of entropy increase, then the entropy of an isolated system only increases and never decreases, i.e. $\partial_i S/\partial t \geq 0$, the system will become more and more disorder with time. However, when the second term of entropy decrease is present and greater than the first term of entropy increase, the total entropy change rate of an isolated system is less than zero, i.e. $\partial_i S/\partial t < 0$, hence the self-organizing structure can spontaneously emerge. It can be seen from this that the entropy decrease contributed by internal attractive interaction force just make an isolated system to be able to overcome the law of entropy increase and produce self-organizing structure. Because the internal interaction force in a great many natural systems is neutral, i.e. $\nabla_q f = 0$, only the first term on the right-hand side of formula (7) remains. This is another reason that the traditional law of entropy increase is universal. If there is only a repulsive interaction force inside the system, i.e. $\nabla_q f < 0$, then the second term on the right-hand side of formula (7) is also positive entropy increase. In other words, two forms of entropy increase in stochastic motion and collision will present simultaneously in this system. Stochastic motion happens in the coordinate space, collision occurs in the momentum space. This system is just described by Boltzmann diffusion equation[10,11] (i.e. Boltzmann equation added to a diffusion term). Here we again see that both stochastic motion and repulsive force cause entropy increase of the system, and attractive force leads to entropy decrease of the system.

Now let us apply formula (7) to discuss two topics of real isolated system.

1. The universe is as an isolated system, the interaction force between stars, galaxies in it is the universal gravitation. The kinetic equation describing their motion is the BBGKY equation of the single-particle probability density [equation (2) without diffusion term][16], the interaction potential is universal gravitation potential $f = -Gm/|q - q_1|$. Substituting this potential into the second term on the right-hand side of formula (7), in principle we can obtain its concrete mathematical result. However, because up to now we do not see the concrete form of the joint probability density $f_2(x, x_1, t)$ in galactic dynamics, we can only according to formula (6) or the second term on the right-hand side of formula (7) affirm that the universal gravitation will cause a new entropy decrease to emerge in the universe. Just this new entropy decrease countervails with the law of entropy increase, and makes that the total entropy of the universe no long only increases and never decreases. Hence the previous inference that the universe tends to heat death seams to lack theoretical basis.

2. Suppose two kinds of black and white gas molecules all with same volume and mass mix in an adiabatic container. If there is no interaction force among gas molecules, the only one result after evolution is that two kinds of black and while molecules homogeneously distribute in the whole container. However, when there is attractive interaction force among black molecules or white molecules, considering the role of the entropy decrease on the right-hand side of formula (7), some local region of the container will present higher density of black molecules or white molecules after evolution. Hence an inhomogeneous structure of density distribution produces.

If the system is open, there is inflow or outflow. Integrating both side of nonequilibrium entropy evolution equation (3) over 6 dimensional phase space and substituting formulas (4) and (6) into this equation, we obtain the formula for total entropy change rate of open system as follows

$$\frac{\partial_o S}{\partial t} = -\int (C \times S_v) \times dA + kD\overline{(\nabla_q q_1)^2} + Nk\overline{(\nabla_q f) \times (\nabla_p q_1)\left[\mathbf{1} - e^{-(q_2-q_1)}\right]} \qquad (8)$$

Where $C = C(q,t)$ is average velocity of fluid, $A$ is the surface area of the system, $S_v = \int S_{vp} dp$ is the entropy density per unit volume. Compared with formula (7), formula (8) is added to a entropy flow term. It shows that the total entropy change rate of open system (on the left-hand side) is determined together by entropy increase (the second term on the right-hand side), entropy decrease (the third term on the right-hand side) and entropy flow (the first term on the right-hand side). The entropy can flow out to environment from the system due to the presence of flow term. As a result, the self-organizing structure can emerge from the system only if the sum of entropy outflow and entropy decrease is greater than entropy increase. It can be seen that for self-organizing structure emerging in open system the entropy decrease caused by internal attractive interaction still play promoting role although it no long must be a dominant power. Here we again see that the power of new entropy decrease can help open system to resist the low of entropy increase and to produce self-organizing structure. Of course, if the internal interaction force is repulsive but not attractive, the third term on the right-hand side of formula (8) is also positive entropy increase. Under such circumstances, if open system want to produce self-organizing structure, it need more entropy flowing out from the system. Hence it is more difficult. If the entropy outflow is so less that the sum of it and entropy decrease is much less than entropy increase, then the total entropy of open system still increase with time.

From formulas (7) (8) it can be seen that on the one hand the entropy in nonequilibrium system always increases (produces), which shows that the power of thermodynamic degradation is eternal; and on the other hand, in addition to the entropy outflow to environment from the system a new entropy decrease can spontaneously emerge in nonequilibrium system with internal attractive interaction force, which manifest the self-organizing power is accumulating strength for showing off. Both of them coexist in a system and theoretical formula and countervail each other.

In view of the importance of nonequilibrium entropy evolution equation (3), here let us give it a brief review. Entropy increase, the third term on the right-hand side, it give a concise statistical formula for the traditional low of entropy increase, i.e. formula (4). Entropy diffusion, the second term on the right-hand side, its presence make us to think that the process of tending to equilibrium is accomplished by entropy diffusing from its high density region to low density region and finally the distribution of entropy density in all the system reaches to uniformity and the entropy of the system approaches to maximum. Hence the essence of tending to equilibrium puzzling physicist for a long time now becomes clear at a glance. Entropy change contributed by interaction potential energy, the fourth term on the right-hand side, it reveals that a new entropy decrease can spontaneously emerge in a nonequilibrium system with internal attractive interaction, and its quantitative expression is formula (6). Obviously, every term among these three terms has itself important physical meaning. As to entropy drift, i.e. entropy flow of open system, the first term on the right-hand side, it is well known in existing theory. If we say, formulas (7) and (8) have united thermodynamic degradation for destroying the order structure and self-organizing evolution for producing the order structure, then nonequilibrium entropy evolution equation (3) may be regarded as a basic equation describing the system evolution including the unity of thermodynamic degradation and self-organizing evolution.

It should be pointed out here that from nonequilibrium entropy evolution equation (3) and the formula for entropy increase rate we have obtained their corresponding dynamic information evolution equation and the formula for information dissipation rate[17]. Similarly, from the formula for entropy decrease rate (6) we can obtain its corresponding formula for information production rate as follows

$$R_i(t) = \int l_i(f) d\bm{x} = -\int l(f) d\bm{x} = -Nk\overline{(\nabla_q f) \times (\nabla_p q_1)\left[1 - e^{-(q_2-q_1)}\right]} \tag{9}$$

Where the information density production rate

$$l_i(f) = -l(f) \tag{9a}$$

Now dynamic information evolution equation corresponding to equation (3) is

$$\frac{\partial I_{vp}}{\partial t} = -\nabla_q \times (\bm{v} I_{vp}) + D\nabla_q^2 I_{vp} - \frac{D}{kf_1}\left[(\nabla_q \ln f_1) I_{vp} - \nabla_q I_{vp}\right]^2 + l_i(f) \tag{10}$$

On the basis of this equation it is easy to present the formulas for the time rate of change of total information in isolated system and open system, which are corresponding to the formulas (7) and (8). The detailed discussion will publish in another paper, it is omitted here.

**5. Brief conclusion**

According to the BBGKY diffusion equation of single particle probability density and the definition of nonequilibrium entropy, many years ago we derived a nonlinear evolution equation of nonequilibrium entropy density changing with time-space, and from this we gave a statistical formula for the law of entropy increase. In this paper starting out from this nonequilibrium entropy evolution equation we proved that the attractive interaction force between microscopic particles could contribute nonequilibrium system to create a new entropy decrease, and derived a statistical formula for this entropy decrease rate. When an isolated system with internal attractive interaction is in nonequilibrium state, the time rate of change of its total entropy is equal to the sum of the formula for the law of entropy increase and the formula for the new entropy decrease rate. Hence, its total entropy no more only increases and never decreases, but is able to decease. It shows that just this entropy decrease is a dominant power for that an isolated system including the universe can overcome the law of entropy increase and spontaneously produce self-organizing structure. Compared with isolated system, the time rate of change of total entropy in open system except entropy increase and entropy decrease still contains an entropy outflow. Therefore for spontaneously creating self-organizing structure in open system the new entropy decrease still can play a promoting role though it is not certainly a dominant power again. Thus, whether isolated system or open system, its total entropy changes contains both thermodynamic degradation of entropy increase on the one hand and self-organizing evolution of entropy decrease on the other hand. Two of them coexist in a theoretical formula and form the unity of opposites.